\documentclass[amsmath,amssymb,floatfix,lengthcheck,superscriptaddress]{revtex4-1}

\usepackage{graphicx}
\usepackage[centering,hmargin=20mm,tmargin=29mm,bmargin=25mm]{geometry}
\usepackage{multirow}
\usepackage{newtxtext}
\usepackage[cmintegrals]{newtxmath} 
\usepackage{bm}
\usepackage{bbm}
\usepackage{etoolbox}
\usepackage{setspace}
\usepackage{subfigure}
\usepackage{enumitem}
\setlist[enumerate]{noitemsep}
\usepackage{titlesec}
\usepackage{booktabs}

\titleformat{\section}{\raggedright\bfseries}{\arabic{section}.}{1em}{}

\usepackage{xcolor}
\usepackage{hyperref}
\usepackage{cleveref}
\hypersetup{colorlinks,
	linkcolor={blue!75!black!80!yellow},
	citecolor={blue!75!black!80!yellow},
	urlcolor={blue!75!black!80!yellow}
}

\makeatletter
\renewcommand\@make@capt@title[2]{	\@ifx@empty\float@link{\@firstofone}{\expandafter\href\expandafter{\float@link}}	\sffamily{\textbf{#1}}\@caption@fignum@sep#2
}
\makeatother

\crefname{Fig}{Fig.}{Figs.}
\Crefname{Fig}{Figure.}{Figures.}

\begin{document}

\title{Generalized Electron Hydrodynamics, Vorticity Coupling, and Hall Viscosity in Crystals}
\author{Georgios Varnavides}
\thanks{These authors contributed equally to this work}
\affiliation{Harvard John A. Paulson School of Engineering and Applied Sciences, Harvard University, Cambridge, MA,USA}
\affiliation{Department of Materials Science and Engineering,
Massachusetts Institute of Technology, Cambridge, MA, USA}
\affiliation{Research Laboratory of Electronics, Massachusetts Institute of Technology, Cambridge, MA, USA}
\author{Adam S. Jermyn}
\thanks{These authors contributed equally to this work}
\affiliation{Center for Computational Astrophysics, Flatiron Institute, New York, NY 10010, USA}
\author{Polina Anikeeva}
\affiliation{Department of Materials Science and Engineering,
Massachusetts Institute of Technology, Cambridge, MA, USA}
\affiliation{Research Laboratory of Electronics, Massachusetts Institute of Technology, Cambridge, MA, USA}
\author{Claudia Felser}
\affiliation{Max-Planck-Institut f\"ur Chemische Physik fester Stoffe, Dresden, Germany}
\author{Prineha Narang}
\email[Electronic address:\;]{prineha@seas.harvard.edu}
\affiliation{Harvard John A. Paulson School of Engineering and Applied Sciences, Harvard University, Cambridge, MA,USA}
\date{\today}

\maketitle
\noindent
\textbf{Theoretical and experimental studies have revealed that electrons in condensed matter can behave hydrodynamically, exhibiting fluid phenomena such as Stokes flow and vortices~\cite{Andreev2011,Levitov2016,Scaffidi2017,Lucas2018,Link2018,Cook2019,Holder2019,Holder2019a,Sulpizio2019}.
Unlike classical fluids, preferred directions inside crystals lift isotropic restrictions, necessitating a generalized treatment of electron hydrodynamics.
We explore electron fluid behaviors arising from the most general viscosity tensors in two and three dimensions, constrained only by thermodynamics and crystal symmetries.
Hexagonal 2D materials such as graphene support flows indistinguishable from those of an isotropic fluid.
By contrast 3D materials including Weyl semimetals~\cite{Gooth2018,PhysRevB.98.115130}, exhibit significant deviations from isotropy.
Breaking time-reversal symmetry, for example in magnetic topological materials, introduces a non-dissipative Hall component to the viscosity tensor~\cite{Avron1998,Banerjee2017, 2019Sci...364..162B,Epstein2019,Holder2019}.
While this vanishes by isotropy in 3D, anisotropic materials can exhibit nonzero Hall viscosity components.
We show that in 3D anisotropic materials the electronic fluid stress can couple to the vorticity without breaking time-reversal symmetry.
Our work demonstrates the anomalous landscape for electron hydrodynamics in systems beyond graphene, and presents experimental geometries to quantify the effects of electronic viscosity.}
\newline
\newline
\noindent
Electron hydrodynamics is observed when microscopic scattering processes conserve momentum over time- and length-scales which are large compared to those of the experimental probe.
However, even as momentum is conserved, free energy may be dissipated from the electronic system, giving rise to a measurable viscosity in the electron flow~\cite{Molenkamp1994,Jong1995,Bandurin2016,Crossno2016,Gooth2018,Ku2019,Moll2016}. 
When momentum is conserved, a fluid obeys Cauchy's laws of motion~\cite{Stokes1966}:
\begin{align}
	\rho \dot{u_i} &=\partial_j \tau_{ji} + \rho f_i \label{eq:cauchy1} \\
	\rho \dot{\sigma_i} &= \partial_j m_{ji} + \rho l_i + \epsilon_{ijk} \tau_{jk} \label{eq:cauchy2}, 
\end{align}
where $\boldsymbol{u}$ and $\rho$ are the fluid velocity and density, $\boldsymbol{f}$ and $\boldsymbol{l}$ are body forces and couples, $\boldsymbol{\tau}$ and $\boldsymbol{m}$ are the fluid stress and couple stress, and $\boldsymbol{\sigma}$ is the intrinsic angular momentum density (internal spin).
The superscript dot denotes the material derivative, $\dot{x} = \partial_t x + u_j \partial_j x$, and $\boldsymbol{\epsilon}$ is the rank-3 alternating tensor. 
We assume couple stresses and body couples to be zero.
In steady state and at experimentally accessible Reynolds numbers~\cite{Ku2019,Mendoza2013}, this implies that the stress tensor is symmetric~\cite{Stokes1966}.
In this limit, electron fluids obey the Navier Stokes equation
\begin{align}
	\rho u_j \partial_j u_i = -\partial_i p + \partial_j \tau_{ji},
	\label{eq:ns}
\end{align}
where $\boldsymbol{\tau}$ is symmetric.
Note that in electron fluids, current density is analogous to the fluid velocity and voltage drops are analogous to changes in pressure.
Assuming the fluid velocity is much smaller than the electronic speed of sound, $\boldsymbol{u}\ll c_s$, the electron fluids are nearly incompressible, thus
\begin{align}
	\partial_i u_i = 0.
	\label{eq:incomp}
\end{align}
In this limit $\rho$ is a constant, which we take to be unity.
Since the fluid stress appears in a divergence, it is defined only up to a constant, which we choose to make $\boldsymbol{\tau}$ vanish when $\boldsymbol{u}$ is uniformly zero~\cite{1959flme.book.....L,1958PhRv..109.1486S}.
We further assume that the fluid stress vanishes for uniform flow, so that it is only a function of the velocity gradient.
Without further loss of generality, the constitutive relation is written to first order as~\cite{1959flme.book.....L}
\begin{align}
	\tau_{ij} = A_{ijkl} \partial_l u_k, 
	\label{eq:constitutive}
\end{align}
where $\boldsymbol{A}$ is the fluid viscosity, a rank-4 tensor relating the fluid velocity gradient ($\partial_j u_i$) and the fluid stress.
Since we take $\boldsymbol{\tau}$ to be symmetric, $\boldsymbol{A}$ is invariant under permutation of its first two indices, i.e. $A_{ijkl}=A_{jikl}$~\cite{1959flme.book.....L,1958PhRv..109.1486S}. 
Viscosity is represented as the sum of three rank-4 tensors basis elements~\cite{Epstein2019}, summarized in~\Cref{table:1}
\begin{align}
	A_{(ij)kl} = \alpha_{((ij)(kl))} + \beta_{[(ij)(kl)]} + \gamma_{(ij)[kl]}. 
	\label{eq:viscosity-decomposition}
\end{align}
Tensor $\boldsymbol{\alpha}$ describes dissipative behavior respecting both stress symmetry and objectivity, i.e. $\alpha_{ijkl}=\alpha_{jikl}=\alpha_{klij}$.
Tensor $\boldsymbol{\beta}$ on the other hand, describes non-dissipative Hall viscosity~\cite{Avron1998,Banerjee2017, 2019Sci...364..162B,Epstein2019,Holder2019}, i.e. $\beta_{ijkl}=-\beta_{klij}$, and is non-zero only when time-reversal symmetry is broken.
Finally, $\boldsymbol{\gamma}$ breaks stress objectivity, i.e. $\gamma_{ijkl}=-\gamma_{ijlk}$, coupling fluid stress to the vorticity.
The fifth column in~\Cref{table:1} specifies whether the tensor is defined according to a handedness convention.
\begin{table}
\centering
\begin{tabular}{l c cccc cc}
\toprule
&
\multicolumn{1}{c}{Tensor} &
\multicolumn{4}{c}{Tensor Symmetries} &
\multicolumn{2}{c}{Indep. Comp.}\\
\cmidrule(lr){3-6} \cmidrule(lr){7-8} 
& & $i\leftrightarrow j$ & $k\leftrightarrow l$ & $ij\leftrightarrow kl$ & type & 3D & 2D \\
\midrule
& $\alpha_{((ij)(kl))}$ & + & + & + & proper & 21 & 6 \\
& $\beta_{[(ij)(kl)]}$ & + & + & - & pseudo & 15 & 3 \\
& $\gamma_{(ij)[kl]}$ & + & - & N/A & pseudo & 18 & 3 \\
\bottomrule
\end{tabular}
\caption{Rank-4 tensors used as orthogonal basis elements for the viscosity tensor.
Even and odd symmetries are represented using parentheses and square brackets respectively.
The fifth column specifies whether the tensor changes sign under mirror operations.}
\label{table:1}
\end{table}

\noindent
In classical fluids the added consideration of rotational invariance requires $\boldsymbol{A}$ to be isotropic, reducing it to the form
\begin{align}
	A_{ijkl} = \lambda \delta_{ij}\delta_{kl} &+ \mu \left( \delta_{il}\delta_{jk} + \delta_{jl}\delta_{ik}\right) \nonumber \\
	&+ \mathcal{B}_1 \left(\epsilon_{ik}\delta_{jl}+\delta_{ik}\epsilon_{jl}\right)+ \Gamma_1 \delta_{ij} \epsilon_{kl} \label{eq:iso},
\end{align}
where $\delta$ is the Kronecker delta, $\boldsymbol{\epsilon}$ is the rank-two alternating tensor, and the Lam\'e parameters $\lambda$ and $\mu$ can be identified as the two independent components of the proper tensor $\boldsymbol{\alpha}$.
In the incompressible case $\lambda$ does not contribute to the stress~\cite{1959flme.book.....L}.
$\mathcal{B}_1$ and $\Gamma_1$ are constants parametrizing terms with the symmetry of $\boldsymbol{\beta}$ and $\boldsymbol{\gamma}$ respectively.
Since $\boldsymbol{\beta}$ and $\boldsymbol{\gamma}$ are pseudo-tensors, the last three terms in~\cref{eq:iso} are only non-zero in two dimensions~\cite{Avron1998,Epstein2019}.
\newline
\newline
\noindent
In crystals, however, there exist preferred directions and we cannot assume rotational invariance.
Instead we must consider the effect of the crystal symmetry given by Neumann's Principle~\cite{Neumann1885,Nye1985}, which requires that physical properties described by rank-4 tensors, such as viscosity, remain invariant under the transformation law:
\begin{align}
	A'_{ijkl} &= |s|^{\eta}s_{im}s_{jn}s_{ko}s_{lp} A_{mnop} \label{eq:neumann},
\end{align}
where $\boldsymbol{s}$ is the space-representation of any given point group symmetry of the crystal, $|s|=\pm 1$ is the determinant of the operation, and $\eta=0$ for proper tensors and $\eta=1$ for pseudo-tensors.
Although~\cref{eq:neumann} relates different components of the viscosity tensor, further constrains must be imposed to ensure the viscosity tensor never does positive work in equation~\eqref{eq:ns}, so that for any velocity field $\boldsymbol{u}$ in $d$-dimensions
\begin{align}
	\int u_i \partial_j (A_{ijkl} \partial_k u_l) d^{d} \boldsymbol{r} \leq 0.
	\label{eq:dissipation}
\end{align}
Letting the Fourier transform of $\boldsymbol{u}$ be
\begin{align}
	\tilde{\boldsymbol{u}}(\boldsymbol{q}) = \int e^{i\boldsymbol{q}\cdot\boldsymbol{r}} u(\boldsymbol{r}) d^{d}\boldsymbol{r}
\end{align}
in $d$-dimensions, we find
\begin{align}
	\int q_j q_k \tilde{u}^*_i(\boldsymbol{q}) \tilde{u}_l(\boldsymbol{q}) A_{ijkl} d^{d} \boldsymbol{q} \geq 0.
\end{align}
This is satisfied when $A_{ijkl}$ has a positive definite biquadratic form in $il$ and $jk$, so we impose this constraint in addition to $ij$ symmetry and crystal symmetry.

\begin{figure*}[ht]
\includegraphics[width=\linewidth]{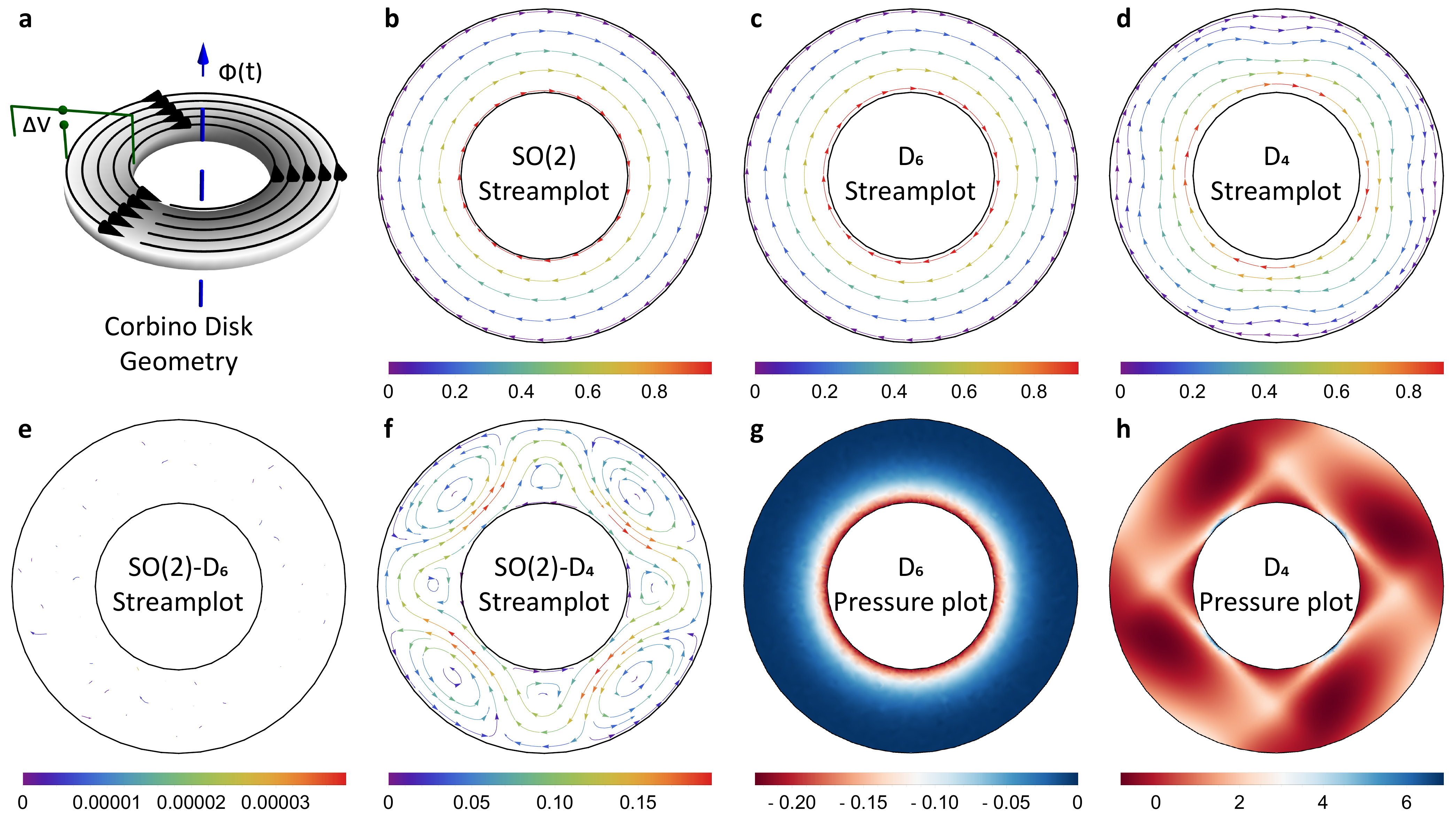}
\caption{\textbf{Effect of viscosity tensor anisotropy on rotational flow in an annulus.}
\textbf{(a)} Corbino disk geometry schematic. The time-varying magnetic flux gives rise to a Lorentz force, inducing rotational electron flow.
Steady state \textbf{(b)} streamplot plot using an isotropic ($SO(2)$) viscosity tensor.
Streamplots using \textbf{(c)} hexagonal ($D_6$), and \textbf{(d)} square ($D_4$) viscosity tensors.
Difference in steady state streamplot between isotropic and \textbf{(e)} $D_6$, and \textbf{(f)} $D_4$ viscosity tensors, highlighting the emergence of steady-state vortices.
Steady state pressure plot using \textbf{(g)} $D_6$ and \textbf{(h)} $D_4$ viscosity tensors, illustrating the breaking of azimuthal symmetry in the latter.}
\label[Fig.]{fig:1}
\end{figure*}

Viscosity tensors are then generated to satisfy the aforementioned constraints~\footnote{The viscosity tensors used in each case are provided in Supplementary Material.}.
The viscosity tensor is assumed to be spatially uniform in all cases.
To demonstrate the differences between these general viscosity tensors and those more strongly constrained by symmetry we solve for the velocity and pressure of low Reynolds number flows in several geometries.
The parametrization of the viscosity tensor in~\cref{eq:viscosity-decomposition} allows us to explore the effects of breaking stress objectivity and time-reversal symmetry.
We highlight the effects  of symmetry in the last two indices ($kl$) because it implies that the stress only couples to the strain rate ($\partial_k u_l + \partial_l u_k$) and not to the vorticity ($\partial_k u_l - \partial_l u_k$).
This is a property of classical fluids, which means that rigid-rotational flows are stress-free, and hence are only sensitive to rotation via weaker effects like the Coriolis force.
Below we demonstrate that with more general viscosity tensors this is not the case, and that the resulting rotational stresses can be probed in experimentally accessible geometries.
\newline
\newline
\noindent
We first consider rotational flow in an annulus with inner radius $R_{\rm inner}=1$ and outer radius $R_{\rm outer}=2$ (Fig.~\ref{fig:1}).
We apply a no-slip condition to the outer boundary, allow the inner boundary to rotate with unit angular velocity $\omega=1$, and solve for the steady state flow at Reynolds number
\begin{align}
	\mathrm{Re} \equiv \frac{\omega R^2_{\rm inner}}{\left|A\right|} = 0.3,
\end{align}
where
\begin{align}
	\left|A\right|^2 = A_{ijkl}A_{ijkl}.
\end{align}
The zero-pressure point is fixed at the bottom of the annulus.
Experimentally, such rotational flows can be achieved by threading a time-varying magnetic flux through a Corbino disk geometry~\cite{Tomadin2014}, shown in Figure~\ref{fig:1}(a).
For a fluid with an isotropic viscosity, the steady-state velocity field rotates rigidly with the angular velocity set by the inner boundary condition (Fig.~\ref{fig:1}(b)).

To investigate the effects of anisotropy in two-dimensional materials, we considered materials with $D_6$ (hexagonal) and $D_4$ (square) symmetry.
Notably, $D_6$ materials do not deviate from isotropic behavior (Fig.~\ref{fig:1}(c)), consistent with experimental observations for graphene~\cite{Sulpizio2019,Ku2019}.
We note that 2D materials with $C_3$ (three-fold), $C_6$ (six-fold), and $D_3$ (triangular) symmetry also exhibit isotropic viscosity tensors (see Supplementary Material).
By contrast, the flow deviates considerably from isotropic behavior in $D_4$ materials (Fig.~\ref{fig:1}(d)).
Figures~\ref{fig:1}(e,f) illustrate these points further, by showing the steady-state velocity flow difference between the isotropic case and $D_6$ and $D_4$ materials respectively.
In the latter, we observe steady-state vortices emerging at $\sim 15\%$ of the bulk flow rate overlaid onto the isotropic velocity field.
While the steady-state pressure field in $D_6$ materials mirrors that of an isotropic fluid (Fig.~\ref{fig:1}(g)), the pressure field in $D_4$ materials also exhibits four vortices (Fig.~\ref{fig:1}(h)), with orientation set by the underlying crystal axes.

\begin{figure*}[ht]
\includegraphics[width=\linewidth]{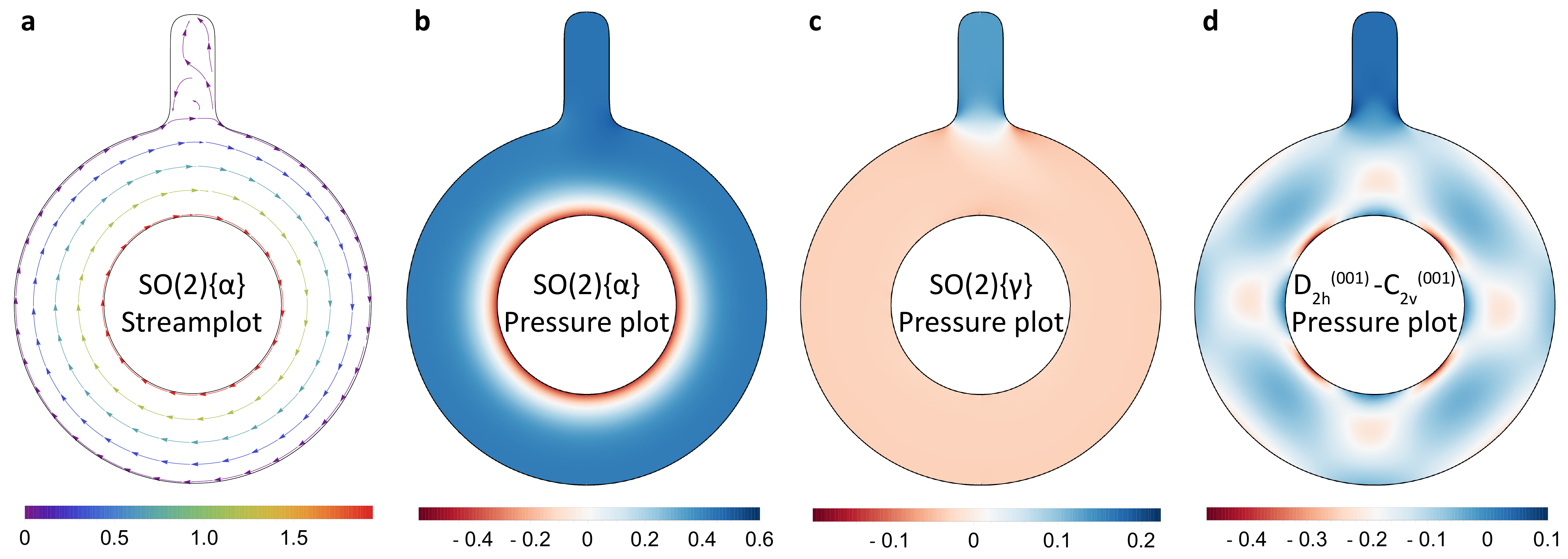}
\caption{\textbf{Proposed setup to quantify effect of viscosity tensor asymmetry and Hall coefficient.}
Steady state \textbf{(a)} streamplot  and \textbf{(b)} pressure plot using a viscosity tensor.
\textbf{(c)} Difference in steady state pressure between the viscosity tensor and the same with additional stress objectivity breaking terms.
The asymmetry introduces an additional pressure-like contribution, which can be directly measured.
\textbf{(d)} Difference in steady state pressure between the $D_{2h}$ and $C_{2v}$ viscosity tensors along the $ab$ plane.}
\label[Fig.]{fig:2}
\end{figure*}

\noindent
We next examine the importance of symmetry in the last two indices of the viscosity tensor.
We calculate the flow profile for the annulus in Figure~\ref{fig:1} scaled by a factor of two, equipped with a pressure gauge, as shown in Figure~\ref{fig:2}(a).
The pressure gauge is a channel with no-slip boundary conditions, allowing us to measure the difference between the flow and a nearly-stationary fluid. 
To isolate the effects of $\mathcal{B}_1$ and $\Gamma_1$ in~\cref{eq:iso}, Figures~\ref{fig:2}(a,b) show the flow and pressure fields in the annulus for a material with isotropic viscosity tensor where $\mathcal{B}_1$ and $\Gamma_1$ have both been set to zero ($SO(2)\{\alpha\}$).
These are nearly unchanged inside the annulus as compared to Figures~\ref{fig:1}(b,g), with a constant pressure in the gauge.
Allowing for non-zero stress-breaking components, i.e. using a material with isotropic viscosity for $\mathcal{B}_1=0$ and $\Gamma_1=0.25$ ($SO(2)\{\gamma\}$), we observe a significant pressure build-up near the gauge.
This is due to the shear stress between the rotating and stationary fluids, while the pressure within the gauge itself is nearly uniform, as shown in Figure~\ref{fig:2}(c).

To quantify the pressure difference between $SO(2)\{\alpha\}$ and $SO(2)\{\gamma\}$, note that the pressure is fixed to zero at a point $p$, the bottom of the annulus domain.
The pressure in the gauge may be written as the path integral
\begin{align}
	p_{\rm gauge} = \int_{p}^{g} \nabla p \cdot d\boldsymbol{s},
\end{align}
where $g$ is a point in the gauge.
At low Reynolds numbers we may neglect $u_j \partial_j u_i$ in equation~\cref{eq:ns}, to find in steady state
\begin{align}
	\nabla p &= A_{ijkl}\partial_i \partial_k u_l. \\
	p_{\rm gauge} &= \int_{p}^{g} A_{ijkl}\partial_i \partial_k u_l d s_j.
\end{align}
Taking into account~\cref{eq:iso} and noting that the changes in fluid flow are negligible, we find:
\begin{align}
	\Delta p_{\rm gauge} = \int_{p}^{g} \Delta A_{ijkl}\partial_i \partial_k u_l d s_j = \Gamma_1 \int_{p}^{g} \partial_i \omega d s_i = \Gamma_1 \Delta \omega,
\end{align}
where $\omega_i=\epsilon_{ijk}\partial_j u_k$ is the vorticity of the flow.
For the geometry used, we find $\Delta p_{\rm gauge} = 0.15$, vorticity in the gauge is zero and that in the annulus is 0.6, so $\Delta \omega=0.6$ (Fig.~\ref{fig:2}(c)).
Since we chose $\Gamma_1=0.25$, we see that in this setup the pressure gauge ($\Delta p_{\rm gauge} = \Gamma_1 \Delta \omega = 0.25 \times 0.6 = 0.15$) is directly sensitive to the asymmetry in the viscosity, which couples the rotation directly to the pressure field and the stress.
We note that the same setup is sensitive to Hall viscosity coefficients for time-reversal broken systems, i.e. for the case where both $\mathcal{B}_1$ and $\Gamma_1$ are non-zero the pressure gauge generalizes to
\begin{align}
	\Delta p_{\rm gauge} = \int_{p}^{g} \Delta A_{ijkl}\partial_i \partial_k u_l d s_j = (\mathcal{B}_1+\Gamma_1) \Delta \omega.
\end{align}
While time reversal and stress objectivity breaking terms persist in two-dimensional isotropic materials, the handedness of the pseudo-tensor implies that mirror operations set them to zero in 3D.
This can be directly observed in comparing low- and high-symmetry three-dimensional crystals.
We consider the same rotational flow along the $ab$ crystal plane of orthorhombic materials, such as the hydrodynamically reported Weyl semi-metal WP\textsubscript{2}~\cite{Gooth2018,PhysRevB.98.115130}.
Along this plane, the difference between the two viscosity tensors can be parametrized as follows:
\begin{align}
	&A^{C_{2v}^{(001)}}_{ijkl} = A^{D_{2h}^{(001)}}_{ijkl}+ \Gamma_2 \delta_{ij}\epsilon_{kl} +  \Gamma_3 \sigma_{ij}^z\epsilon_{kl} \nonumber \\
	&\qquad + \mathcal{B}_2 \left(\delta_{li}\epsilon_{jk}-\epsilon_{li}\delta_{jk}\right) + \mathcal{B}_3 \left(\delta_{ij}\sigma_{kl}^x-\sigma_{ij}^x\delta_{kl}\right) ,
	\label{eq:C2vD2h}
\end{align}
where $\mathcal{B}_2$, $\mathcal{B}_3$, $\Gamma_2$, and $\Gamma_3$ are constants parametrizing terms with the symmetry of $\boldsymbol{\beta}$ and $\boldsymbol{\gamma}$ respectively, $\boldsymbol{\sigma^x}$ and $\boldsymbol{\sigma^z}$ are Pauli matrices.
Figure~\ref{fig:2}(d) shows the pressure difference between a material with $D_{2h}$ symmetry and one with $C_{2v}$ symmetry (for $\mathcal{B}_2=\mathcal{B}_3=\Gamma_3=0$ and $\Gamma_2=0.25$), indicating the same pressure build-up as in Figure~\ref{fig:2}(c) inside the gauge along with non-trivial pressure structure in the annulus.

\begin{figure*}[ht]
\includegraphics[width=\linewidth]{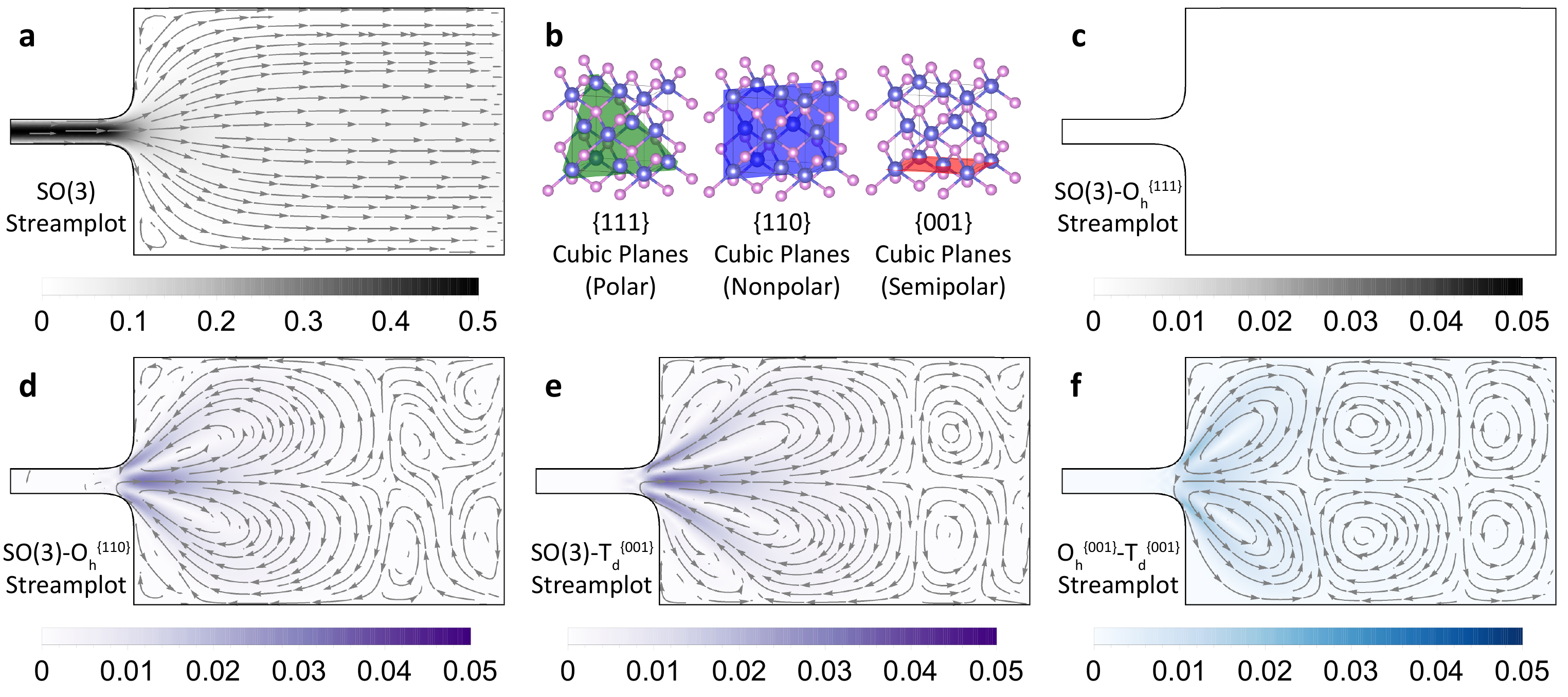}
\caption{\textbf{Three dimensional projected flows through an expanding geometry.}
\textbf{(a)} Steady state streamplot using an isotropic ($SO(3)$) viscosity tensor.
\textbf{(b)} High-symmetry family of planes in cubic crystals.
In crystals with $T_d$ (tetrahedral) symmetry, these are further identified as polar $\{111\}$, nonpolar $\{110\}$, and semipolar $\{001\}$.
Difference in steady state streamplot between using an isotropic viscosity tensor and using \textbf{(c)} (111)-projected, \textbf{(d)} (110)-projected, and \textbf{(e)} (001)-projected viscosity tensors of cubic crystals with $T_{d}$ symmetry.
Note that \textbf{(c)} shows no difference.
\textbf{(f)} Effect of viscosity tensor asymmetry difference between $O_h$ and $T_d$ crystals along a semipolar $\{001\}$ plane.}
\label[Fig.]{fig:3}
\end{figure*}
\noindent
Finally, we consider flow through an expanding channel along high-symmetry planes in 3D.
This geometry has been proposed as a diagnostic of electron hydrodynamics because it naturally generates vortices, not present in ordinary ohmic flow.
The case with isotropic viscosity is shown in Figure~\ref{fig:3}(a), where the small vortices that form in the corners are clearly detached from the bulk of the flow.
We consider the $T_d$ (tetrahedron) and $O_h$ (cubic) point groups.
In particular, we consider flows along the polar $\{111\}$, nonpolar $\{110\}$, and semipolar $\{001\}$ family of planes (Figure~\ref{fig:3}(b)):
\begin{subequations}
\begin{align}
	A^{T_{d}^{(111)}}_{ijkl} &= A^{O_{h}^{(111)}}_{ijkl}+ \mathcal{B}_4\left(\sigma_{ij}^x\sigma_{kl}^z-\sigma_{ij}^z\sigma_{kl}^x\right)+ \Gamma_4 \delta_{ij}\epsilon_{kl} \label{eq:TdOh111}\\
	A^{T_{d}^{(110)}}_{ijkl} &= A^{O_{h}^{(110)}}_{ijkl} \label{eq:TdOh110} \\
	A^{T_{d}^{(001)}}_{ijkl} &= A^{O_{h}^{(001)}}_{ijkl} + \mathcal{B}_5\left(\sigma_{ij}^z\delta_{kl}^z-\delta_{ij}\sigma_{kl}^z\right)+ \Gamma_5 \sigma_{ij}^x\epsilon_{kl} \label{eq:TdOh001}.
\end{align}
\end{subequations}
Along these planes, the difference between the two viscosity tensors can be parametrized according to~\cref{eq:TdOh111,eq:TdOh110,eq:TdOh001}.
We impose fully-developed (parabolic) inlet and outlet flows with constant discharge, and solve for the steady state flow at low Reynolds number.
Figure~\ref{fig:3}(c) shows the difference between the flow in an isotropic material and the flow in a cubic material along a $\{111\}$ close-packed plane, which exhibits rotational invariance.
Along the nonpolar $\{110\}$ planes, terms with $\boldsymbol{\beta}$ and $\boldsymbol{\gamma}$ symmetry vanish.
However, $\boldsymbol{A^{O_{h}^{(110)}}}$ is anisotropic along this plane, with Figure~\ref{fig:3}(d) showing the difference in flow between the isotropic case.
Finally, along the semipolar $\{001\}$ family of planes, the viscosity tensor is both anisotropic (Fig.~\ref{fig:3}(e)), and asymmetric.
Figure~\ref{fig:3}(f) quantifies the additional vortices generated by the asymmetry at $\sim 10\%$, for $ \mathcal{B}_5=0$ and $\Gamma_5=0.25$.
\newline
\newline
\noindent
We found that electron fluids in crystals with anisotropic and asymmetric viscosity tensors can exhibit steady-state fluid behaviors not observed in classical fluids.
Even a minor deviation from isotropy allows the fluid stress to couple to the fluid vorticity with or without breaking time-reversal symmetry, for the case of Hall viscosity and objectivity-breaking viscosity respectively.
Recent measurements of spatially-resolved flows~\cite{Ku2019,Ella2019,Sulpizio2019}, suggest that these effects can be directly observed in systems beyond graphene.
Our findings further hint at potential applications.
For instance, the pressure gauge in Figure~\ref{fig:2} could be used as a magnetometer, converting a time-varying magnetic flux through a modified corbino disk geometry into current in the annulus, and ultimately into a voltage drop between it and the gauge.
Our work highlights the importance of crystal symmetry on electronic flow, and invites further exploration of time-dependent flows in systems with internal spin degrees of freedom and asymmetric stress tensors.
\newline
\newline
\noindent
\textbf{\textit{Acknowledgments.}}
The authors thank Prof. Andrew Lucas of the University of Colorado Boulder for fruitful discussions.
The authors acknowledge funding from the Defense Advanced Research Projects Agency (DARPA) Defense Sciences Office (DSO) Driven and Nonequilibrium Quantum Systems program.
ASJ is supported by the Flatiron Institute of the Simons Foundation. 
PN is a Moore Inventor Fellow supported by the Gordon and Betty Moore Foundation.
\newline
\newline
\noindent
\textbf{\textit{Author Contributions.}}
GV and ASJ jointly conceived the ideas and developed the framework.
GV, CF, and PN identified material systems and link with transport measurements of topological systems.
GV, PA, and PN jointly worked on introducing the effects of crystal symmetries.
All authors discussed the findings and contributed to the writing of the manuscript.
\newline
\newline
\noindent
\textbf{\textit{Competing Financial Interests.}}
The authors have no competing financial interests.
 

\begin{thebibliography}{30}\makeatletter
\providecommand \@ifxundefined [1]{ \@ifx{#1\undefined}
}\providecommand \@ifnum [1]{ \ifnum #1\expandafter \@firstoftwo
 \else \expandafter \@secondoftwo
 \fi
}\providecommand \@ifx [1]{ \ifx #1\expandafter \@firstoftwo
 \else \expandafter \@secondoftwo
 \fi
}\providecommand \natexlab [1]{#1}\providecommand \enquote  [1]{``#1''}\providecommand \bibnamefont  [1]{#1}\providecommand \bibfnamefont [1]{#1}\providecommand \citenamefont [1]{#1}\providecommand \href@noop [0]{\@secondoftwo}\providecommand \href [0]{\begingroup \@sanitize@url \@href}\providecommand \@href[1]{\@@startlink{#1}\@@href}\providecommand \@@href[1]{\endgroup#1\@@endlink}\providecommand \@sanitize@url [0]{\catcode `\\12\catcode `\$12\catcode
  `\&12\catcode `\#12\catcode `\^12\catcode `\_12\catcode `\%12\relax}\providecommand \@@startlink[1]{}\providecommand \@@endlink[0]{}\providecommand \url  [0]{\begingroup\@sanitize@url \@url }\providecommand \@url [1]{\endgroup\@href {#1}{\urlprefix }}\providecommand \urlprefix  [0]{URL }\providecommand \Eprint [0]{\href }\providecommand \doibase [0]{http://dx.doi.org/}\providecommand \selectlanguage [0]{\@gobble}\providecommand \bibinfo  [0]{\@secondoftwo}\providecommand \bibfield  [0]{\@secondoftwo}\providecommand \translation [1]{[#1]}\providecommand \BibitemOpen [0]{}\providecommand \bibitemStop [0]{}\providecommand \bibitemNoStop [0]{.\EOS\space}\providecommand \EOS [0]{\spacefactor3000\relax}\providecommand \BibitemShut  [1]{\csname bibitem#1\endcsname}\let\auto@bib@innerbib\@empty
\bibitem [{\citenamefont {Andreev}\ \emph {et~al.}(2011)\citenamefont
  {Andreev}, \citenamefont {Kivelson},\ and\ \citenamefont
  {Spivak}}]{Andreev2011}  \BibitemOpen
  \bibfield  {author} {\bibinfo {author} {\bibfnamefont {A.~V.}\ \bibnamefont
  {Andreev}}, \bibinfo {author} {\bibfnamefont {S.~A.}\ \bibnamefont
  {Kivelson}}, \ and\ \bibinfo {author} {\bibfnamefont {B.}~\bibnamefont
  {Spivak}},\ }\href {\doibase 10.1103/physrevlett.106.256804} {\bibfield
  {journal} {\bibinfo  {journal} {Physical Review Letters}\ }\textbf {\bibinfo
  {volume} {106}} (\bibinfo {year} {2011}),\
  10.1103/physrevlett.106.256804}\BibitemShut {NoStop}\bibitem [{\citenamefont {Levitov}\ and\ \citenamefont
  {Falkovich}(2016)}]{Levitov2016}  \BibitemOpen
  \bibfield  {author} {\bibinfo {author} {\bibfnamefont {L.}~\bibnamefont
  {Levitov}}\ and\ \bibinfo {author} {\bibfnamefont {G.}~\bibnamefont
  {Falkovich}},\ }\href {\doibase 10.1038/nphys3667} {\bibfield  {journal}
  {\bibinfo  {journal} {Nature Physics}\ }\textbf {\bibinfo {volume} {12}},\
  \bibinfo {pages} {672} (\bibinfo {year} {2016})}\BibitemShut {NoStop}\bibitem [{\citenamefont {Scaffidi}\ \emph {et~al.}(2017)\citenamefont
  {Scaffidi}, \citenamefont {Nandi}, \citenamefont {Schmidt}, \citenamefont
  {Mackenzie},\ and\ \citenamefont {Moore}}]{Scaffidi2017}  \BibitemOpen
  \bibfield  {author} {\bibinfo {author} {\bibfnamefont {T.}~\bibnamefont
  {Scaffidi}}, \bibinfo {author} {\bibfnamefont {N.}~\bibnamefont {Nandi}},
  \bibinfo {author} {\bibfnamefont {B.}~\bibnamefont {Schmidt}}, \bibinfo
  {author} {\bibfnamefont {A.~P.}\ \bibnamefont {Mackenzie}}, \ and\ \bibinfo
  {author} {\bibfnamefont {J.~E.}\ \bibnamefont {Moore}},\ }\href {\doibase
  10.1103/physrevlett.118.226601} {\bibfield  {journal} {\bibinfo  {journal}
  {Physical Review Letters}\ }\textbf {\bibinfo {volume} {118}} (\bibinfo
  {year} {2017}),\ 10.1103/physrevlett.118.226601}\BibitemShut {NoStop}\bibitem [{\citenamefont {Lucas}\ and\ \citenamefont {Fong}(2018)}]{Lucas2018}  \BibitemOpen
  \bibfield  {author} {\bibinfo {author} {\bibfnamefont {A.}~\bibnamefont
  {Lucas}}\ and\ \bibinfo {author} {\bibfnamefont {K.~C.}\ \bibnamefont
  {Fong}},\ }\href {\doibase 10.1088/1361-648x/aaa274} {\bibfield  {journal}
  {\bibinfo  {journal} {Journal of Physics: Condensed Matter}\ }\textbf
  {\bibinfo {volume} {30}},\ \bibinfo {pages} {053001} (\bibinfo {year}
  {2018})}\BibitemShut {NoStop}\bibitem [{\citenamefont {Link}\ \emph {et~al.}(2018)\citenamefont {Link},
  \citenamefont {Narozhny}, \citenamefont {Kiselev},\ and\ \citenamefont
  {Schmalian}}]{Link2018}  \BibitemOpen
  \bibfield  {author} {\bibinfo {author} {\bibfnamefont {J.~M.}\ \bibnamefont
  {Link}}, \bibinfo {author} {\bibfnamefont {B.~N.}\ \bibnamefont {Narozhny}},
  \bibinfo {author} {\bibfnamefont {E.~I.}\ \bibnamefont {Kiselev}}, \ and\
  \bibinfo {author} {\bibfnamefont {J.}~\bibnamefont {Schmalian}},\ }\href
  {\doibase 10.1103/physrevlett.120.196801} {\bibfield  {journal} {\bibinfo
  {journal} {Physical Review Letters}\ }\textbf {\bibinfo {volume} {120}}
  (\bibinfo {year} {2018}),\ 10.1103/physrevlett.120.196801}\BibitemShut
  {NoStop}\bibitem [{\citenamefont {Cook}\ and\ \citenamefont {Lucas}(2019)}]{Cook2019}  \BibitemOpen
  \bibfield  {author} {\bibinfo {author} {\bibfnamefont {C.~Q.}\ \bibnamefont
  {Cook}}\ and\ \bibinfo {author} {\bibfnamefont {A.}~\bibnamefont {Lucas}},\
  }\href {\doibase 10.1103/physrevb.99.235148} {\bibfield  {journal} {\bibinfo
  {journal} {Physical Review B}\ }\textbf {\bibinfo {volume} {99}} (\bibinfo
  {year} {2019}),\ 10.1103/physrevb.99.235148}\BibitemShut {NoStop}\bibitem [{\citenamefont {Holder}\ \emph
  {et~al.}(2019{\natexlab{a}})\citenamefont {Holder}, \citenamefont {Queiroz},\
  and\ \citenamefont {Stern}}]{Holder2019}  \BibitemOpen
  \bibfield  {author} {\bibinfo {author} {\bibfnamefont {T.}~\bibnamefont
  {Holder}}, \bibinfo {author} {\bibfnamefont {R.}~\bibnamefont {Queiroz}}, \
  and\ \bibinfo {author} {\bibfnamefont {A.}~\bibnamefont {Stern}},\ }\href
  {\doibase 10.1103/physrevlett.123.106801} {\bibfield  {journal} {\bibinfo
  {journal} {Physical Review Letters}\ }\textbf {\bibinfo {volume} {123}}
  (\bibinfo {year} {2019}{\natexlab{a}}),\
  10.1103/physrevlett.123.106801}\BibitemShut {NoStop}\bibitem [{\citenamefont {Holder}\ \emph
  {et~al.}(2019{\natexlab{b}})\citenamefont {Holder}, \citenamefont {Queiroz},
  \citenamefont {Scaffidi}, \citenamefont {Silberstein}, \citenamefont {Rozen},
  \citenamefont {Sulpizio}, \citenamefont {Ella}, \citenamefont {Ilani},\ and\
  \citenamefont {Stern}}]{Holder2019a}  \BibitemOpen
  \bibfield  {author} {\bibinfo {author} {\bibfnamefont {T.}~\bibnamefont
  {Holder}}, \bibinfo {author} {\bibfnamefont {R.}~\bibnamefont {Queiroz}},
  \bibinfo {author} {\bibfnamefont {T.}~\bibnamefont {Scaffidi}}, \bibinfo
  {author} {\bibfnamefont {N.}~\bibnamefont {Silberstein}}, \bibinfo {author}
  {\bibfnamefont {A.}~\bibnamefont {Rozen}}, \bibinfo {author} {\bibfnamefont
  {J.~A.}\ \bibnamefont {Sulpizio}}, \bibinfo {author} {\bibfnamefont
  {L.}~\bibnamefont {Ella}}, \bibinfo {author} {\bibfnamefont {S.}~\bibnamefont
  {Ilani}}, \ and\ \bibinfo {author} {\bibfnamefont {A.}~\bibnamefont
  {Stern}},\ }\href {\doibase 10.1103/physrevb.100.245305} {\bibfield
  {journal} {\bibinfo  {journal} {Physical Review B}\ }\textbf {\bibinfo
  {volume} {100}} (\bibinfo {year} {2019}{\natexlab{b}}),\
  10.1103/physrevb.100.245305}\BibitemShut {NoStop}\bibitem [{\citenamefont {Sulpizio}\ \emph {et~al.}(2019)\citenamefont
  {Sulpizio}, \citenamefont {Ella}, \citenamefont {Rozen}, \citenamefont
  {Birkbeck}, \citenamefont {Perello}, \citenamefont {Dutta}, \citenamefont
  {Ben-Shalom}, \citenamefont {Taniguchi}, \citenamefont {Watanabe},
  \citenamefont {Holder}, \citenamefont {Queiroz}, \citenamefont {Principi},
  \citenamefont {Stern}, \citenamefont {Scaffidi}, \citenamefont {Geim},\ and\
  \citenamefont {Ilani}}]{Sulpizio2019}  \BibitemOpen
  \bibfield  {author} {\bibinfo {author} {\bibfnamefont {J.~A.}\ \bibnamefont
  {Sulpizio}}, \bibinfo {author} {\bibfnamefont {L.}~\bibnamefont {Ella}},
  \bibinfo {author} {\bibfnamefont {A.}~\bibnamefont {Rozen}}, \bibinfo
  {author} {\bibfnamefont {J.}~\bibnamefont {Birkbeck}}, \bibinfo {author}
  {\bibfnamefont {D.~J.}\ \bibnamefont {Perello}}, \bibinfo {author}
  {\bibfnamefont {D.}~\bibnamefont {Dutta}}, \bibinfo {author} {\bibfnamefont
  {M.}~\bibnamefont {Ben-Shalom}}, \bibinfo {author} {\bibfnamefont
  {T.}~\bibnamefont {Taniguchi}}, \bibinfo {author} {\bibfnamefont
  {K.}~\bibnamefont {Watanabe}}, \bibinfo {author} {\bibfnamefont
  {T.}~\bibnamefont {Holder}}, \bibinfo {author} {\bibfnamefont
  {R.}~\bibnamefont {Queiroz}}, \bibinfo {author} {\bibfnamefont
  {A.}~\bibnamefont {Principi}}, \bibinfo {author} {\bibfnamefont
  {A.}~\bibnamefont {Stern}}, \bibinfo {author} {\bibfnamefont
  {T.}~\bibnamefont {Scaffidi}}, \bibinfo {author} {\bibfnamefont {A.~K.}\
  \bibnamefont {Geim}}, \ and\ \bibinfo {author} {\bibfnamefont
  {S.}~\bibnamefont {Ilani}},\ }\href {\doibase 10.1038/s41586-019-1788-9}
  {\bibfield  {journal} {\bibinfo  {journal} {Nature}\ }\textbf {\bibinfo
  {volume} {576}},\ \bibinfo {pages} {75} (\bibinfo {year} {2019})}\BibitemShut
  {NoStop}\bibitem [{\citenamefont {Gooth}\ \emph {et~al.}(2018)\citenamefont {Gooth},
  \citenamefont {Menges}, \citenamefont {Kumar}, \citenamefont {Sü$\upbeta$},
  \citenamefont {Shekhar}, \citenamefont {Sun}, \citenamefont {Drechsler},
  \citenamefont {Zierold}, \citenamefont {Felser},\ and\ \citenamefont
  {Gotsmann}}]{Gooth2018}  \BibitemOpen
  \bibfield  {author} {\bibinfo {author} {\bibfnamefont {J.}~\bibnamefont
  {Gooth}}, \bibinfo {author} {\bibfnamefont {F.}~\bibnamefont {Menges}},
  \bibinfo {author} {\bibfnamefont {N.}~\bibnamefont {Kumar}}, \bibinfo
  {author} {\bibfnamefont {V.}~\bibnamefont {Sü$\upbeta$}}, \bibinfo {author}
  {\bibfnamefont {C.}~\bibnamefont {Shekhar}}, \bibinfo {author} {\bibfnamefont
  {Y.}~\bibnamefont {Sun}}, \bibinfo {author} {\bibfnamefont {U.}~\bibnamefont
  {Drechsler}}, \bibinfo {author} {\bibfnamefont {R.}~\bibnamefont {Zierold}},
  \bibinfo {author} {\bibfnamefont {C.}~\bibnamefont {Felser}}, \ and\ \bibinfo
  {author} {\bibfnamefont {B.}~\bibnamefont {Gotsmann}},\ }\href {\doibase
  10.1038/s41467-018-06688-y} {\bibfield  {journal} {\bibinfo  {journal}
  {Nature Communications}\ }\textbf {\bibinfo {volume} {9}} (\bibinfo {year}
  {2018}),\ 10.1038/s41467-018-06688-y}\BibitemShut {NoStop}\bibitem [{\citenamefont {Coulter}\ \emph {et~al.}(2018)\citenamefont
  {Coulter}, \citenamefont {Sundararaman},\ and\ \citenamefont
  {Narang}}]{PhysRevB.98.115130}  \BibitemOpen
  \bibfield  {author} {\bibinfo {author} {\bibfnamefont {J.}~\bibnamefont
  {Coulter}}, \bibinfo {author} {\bibfnamefont {R.}~\bibnamefont
  {Sundararaman}}, \ and\ \bibinfo {author} {\bibfnamefont {P.}~\bibnamefont
  {Narang}},\ }\href {\doibase 10.1103/PhysRevB.98.115130} {\bibfield
  {journal} {\bibinfo  {journal} {Phys. Rev. B}\ }\textbf {\bibinfo {volume}
  {98}},\ \bibinfo {pages} {115130} (\bibinfo {year} {2018})}\BibitemShut
  {NoStop}\bibitem [{\citenamefont {Avron}(1998)}]{Avron1998}  \BibitemOpen
  \bibfield  {author} {\bibinfo {author} {\bibfnamefont {J.~E.}\ \bibnamefont
  {Avron}},\ }\href {\doibase 10.1023/a:1023084404080} {\bibfield  {journal}
  {\bibinfo  {journal} {Journal of Statistical Physics}\ }\textbf {\bibinfo
  {volume} {92}},\ \bibinfo {pages} {543} (\bibinfo {year} {1998})}\BibitemShut
  {NoStop}\bibitem [{\citenamefont {Banerjee}\ \emph {et~al.}(2017)\citenamefont
  {Banerjee}, \citenamefont {Souslov}, \citenamefont {Abanov},\ and\
  \citenamefont {Vitelli}}]{Banerjee2017}  \BibitemOpen
  \bibfield  {author} {\bibinfo {author} {\bibfnamefont {D.}~\bibnamefont
  {Banerjee}}, \bibinfo {author} {\bibfnamefont {A.}~\bibnamefont {Souslov}},
  \bibinfo {author} {\bibfnamefont {A.~G.}\ \bibnamefont {Abanov}}, \ and\
  \bibinfo {author} {\bibfnamefont {V.}~\bibnamefont {Vitelli}},\ }\href
  {\doibase 10.1038/s41467-017-01378-7} {\bibfield  {journal} {\bibinfo
  {journal} {Nature Communications}\ }\textbf {\bibinfo {volume} {8}} (\bibinfo
  {year} {2017}),\ 10.1038/s41467-017-01378-7}\BibitemShut {NoStop}\bibitem [{\citenamefont {{Berdyugin}}\ \emph {et~al.}(2019)\citenamefont
  {{Berdyugin}}, \citenamefont {{Xu}}, \citenamefont {{Pellegrino}},
  \citenamefont {{Krishna Kumar}}, \citenamefont {{Principi}}, \citenamefont
  {{Torre}}, \citenamefont {{Ben Shalom}}, \citenamefont {{Taniguchi}},
  \citenamefont {{Watanabe}}, \citenamefont {{Grigorieva}}, \citenamefont
  {{Polini}}, \citenamefont {{Geim}},\ and\ \citenamefont {{Band
  urin}}}]{2019Sci...364..162B}  \BibitemOpen
  \bibfield  {author} {\bibinfo {author} {\bibfnamefont {A.~I.}\ \bibnamefont
  {{Berdyugin}}}, \bibinfo {author} {\bibfnamefont {S.~G.}\ \bibnamefont
  {{Xu}}}, \bibinfo {author} {\bibfnamefont {F.~M.~D.}\ \bibnamefont
  {{Pellegrino}}}, \bibinfo {author} {\bibfnamefont {R.}~\bibnamefont {{Krishna
  Kumar}}}, \bibinfo {author} {\bibfnamefont {A.}~\bibnamefont {{Principi}}},
  \bibinfo {author} {\bibfnamefont {I.}~\bibnamefont {{Torre}}}, \bibinfo
  {author} {\bibfnamefont {M.}~\bibnamefont {{Ben Shalom}}}, \bibinfo {author}
  {\bibfnamefont {T.}~\bibnamefont {{Taniguchi}}}, \bibinfo {author}
  {\bibfnamefont {K.}~\bibnamefont {{Watanabe}}}, \bibinfo {author}
  {\bibfnamefont {I.~V.}\ \bibnamefont {{Grigorieva}}}, \bibinfo {author}
  {\bibfnamefont {M.}~\bibnamefont {{Polini}}}, \bibinfo {author}
  {\bibfnamefont {A.~K.}\ \bibnamefont {{Geim}}}, \ and\ \bibinfo {author}
  {\bibfnamefont {D.~A.}\ \bibnamefont {{Band urin}}},\ }\href {\doibase
  10.1126/science.aau0685} {\bibfield  {journal} {\bibinfo  {journal}
  {Science}\ }\textbf {\bibinfo {volume} {364}},\ \bibinfo {pages} {162}
  (\bibinfo {year} {2019})},\ \Eprint {http://arxiv.org/abs/1806.01606}
  {arXiv:1806.01606 [cond-mat.mes-hall]} \BibitemShut {NoStop}\bibitem [{\citenamefont {Epstein}\ and\ \citenamefont
  {Mandadapu}()}]{Epstein2019}  \BibitemOpen
  \bibfield  {author} {\bibinfo {author} {\bibfnamefont {J.~M.}\ \bibnamefont
  {Epstein}}\ and\ \bibinfo {author} {\bibfnamefont {K.~K.}\ \bibnamefont
  {Mandadapu}},\ }\href@noop {} {\ }\Eprint
  {http://arxiv.org/abs/http://arxiv.org/abs/1907.10041v1}
  {http://arxiv.org/abs/1907.10041v1} \BibitemShut {NoStop}\bibitem [{\citenamefont {Molenkamp}\ and\ \citenamefont
  {de~Jong}(1994)}]{Molenkamp1994}  \BibitemOpen
  \bibfield  {author} {\bibinfo {author} {\bibfnamefont {L.}~\bibnamefont
  {Molenkamp}}\ and\ \bibinfo {author} {\bibfnamefont {M.}~\bibnamefont
  {de~Jong}},\ }\href {\doibase 10.1016/0038-1101(94)90244-5} {\bibfield
  {journal} {\bibinfo  {journal} {Solid-State Electronics}\ }\textbf {\bibinfo
  {volume} {37}},\ \bibinfo {pages} {551} (\bibinfo {year} {1994})}\BibitemShut
  {NoStop}\bibitem [{\citenamefont {de~Jong}\ and\ \citenamefont
  {Molenkamp}(1995)}]{Jong1995}  \BibitemOpen
  \bibfield  {author} {\bibinfo {author} {\bibfnamefont {M.~J.~M.}\
  \bibnamefont {de~Jong}}\ and\ \bibinfo {author} {\bibfnamefont {L.~W.}\
  \bibnamefont {Molenkamp}},\ }\href {\doibase 10.1103/physrevb.51.13389}
  {\bibfield  {journal} {\bibinfo  {journal} {Physical Review B}\ }\textbf
  {\bibinfo {volume} {51}},\ \bibinfo {pages} {13389} (\bibinfo {year}
  {1995})}\BibitemShut {NoStop}\bibitem [{\citenamefont {Bandurin}\ \emph {et~al.}(2016)\citenamefont
  {Bandurin}, \citenamefont {Torre}, \citenamefont {Kumar}, \citenamefont
  {Shalom}, \citenamefont {Tomadin}, \citenamefont {Principi}, \citenamefont
  {Auton}, \citenamefont {Khestanova}, \citenamefont {Novoselov}, \citenamefont
  {Grigorieva}, \citenamefont {Ponomarenko}, \citenamefont {Geim},\ and\
  \citenamefont {Polini}}]{Bandurin2016}  \BibitemOpen
  \bibfield  {author} {\bibinfo {author} {\bibfnamefont {D.~A.}\ \bibnamefont
  {Bandurin}}, \bibinfo {author} {\bibfnamefont {I.}~\bibnamefont {Torre}},
  \bibinfo {author} {\bibfnamefont {R.~K.}\ \bibnamefont {Kumar}}, \bibinfo
  {author} {\bibfnamefont {M.~B.}\ \bibnamefont {Shalom}}, \bibinfo {author}
  {\bibfnamefont {A.}~\bibnamefont {Tomadin}}, \bibinfo {author} {\bibfnamefont
  {A.}~\bibnamefont {Principi}}, \bibinfo {author} {\bibfnamefont {G.~H.}\
  \bibnamefont {Auton}}, \bibinfo {author} {\bibfnamefont {E.}~\bibnamefont
  {Khestanova}}, \bibinfo {author} {\bibfnamefont {K.~S.}\ \bibnamefont
  {Novoselov}}, \bibinfo {author} {\bibfnamefont {I.~V.}\ \bibnamefont
  {Grigorieva}}, \bibinfo {author} {\bibfnamefont {L.~A.}\ \bibnamefont
  {Ponomarenko}}, \bibinfo {author} {\bibfnamefont {A.~K.}\ \bibnamefont
  {Geim}}, \ and\ \bibinfo {author} {\bibfnamefont {M.}~\bibnamefont
  {Polini}},\ }\href {\doibase 10.1126/science.aad0201} {\bibfield  {journal}
  {\bibinfo  {journal} {Science}\ }\textbf {\bibinfo {volume} {351}},\ \bibinfo
  {pages} {1055} (\bibinfo {year} {2016})}\BibitemShut {NoStop}\bibitem [{\citenamefont {Crossno}\ \emph {et~al.}(2016)\citenamefont
  {Crossno}, \citenamefont {Shi}, \citenamefont {Wang}, \citenamefont {Liu},
  \citenamefont {Harzheim}, \citenamefont {Lucas}, \citenamefont {Sachdev},
  \citenamefont {Kim}, \citenamefont {Taniguchi}, \citenamefont {Watanabe},
  \citenamefont {Ohki},\ and\ \citenamefont {Fong}}]{Crossno2016}  \BibitemOpen
  \bibfield  {author} {\bibinfo {author} {\bibfnamefont {J.}~\bibnamefont
  {Crossno}}, \bibinfo {author} {\bibfnamefont {J.~K.}\ \bibnamefont {Shi}},
  \bibinfo {author} {\bibfnamefont {K.}~\bibnamefont {Wang}}, \bibinfo {author}
  {\bibfnamefont {X.}~\bibnamefont {Liu}}, \bibinfo {author} {\bibfnamefont
  {A.}~\bibnamefont {Harzheim}}, \bibinfo {author} {\bibfnamefont
  {A.}~\bibnamefont {Lucas}}, \bibinfo {author} {\bibfnamefont
  {S.}~\bibnamefont {Sachdev}}, \bibinfo {author} {\bibfnamefont
  {P.}~\bibnamefont {Kim}}, \bibinfo {author} {\bibfnamefont {T.}~\bibnamefont
  {Taniguchi}}, \bibinfo {author} {\bibfnamefont {K.}~\bibnamefont {Watanabe}},
  \bibinfo {author} {\bibfnamefont {T.~A.}\ \bibnamefont {Ohki}}, \ and\
  \bibinfo {author} {\bibfnamefont {K.~C.}\ \bibnamefont {Fong}},\ }\href
  {\doibase 10.1126/science.aad0343} {\bibfield  {journal} {\bibinfo  {journal}
  {Science}\ }\textbf {\bibinfo {volume} {351}},\ \bibinfo {pages} {1058}
  (\bibinfo {year} {2016})}\BibitemShut {NoStop}\bibitem [{\citenamefont {Ku}\ \emph {et~al.}()\citenamefont {Ku},
  \citenamefont {Zhou}, \citenamefont {Li}, \citenamefont {Shin}, \citenamefont
  {Shi}, \citenamefont {Burch}, \citenamefont {Zhang}, \citenamefont {Casola},
  \citenamefont {Taniguchi}, \citenamefont {Watanabe}, \citenamefont {Kim},
  \citenamefont {Yacoby},\ and\ \citenamefont {Walsworth}}]{Ku2019}  \BibitemOpen
  \bibfield  {author} {\bibinfo {author} {\bibfnamefont {M.~J.~H.}\
  \bibnamefont {Ku}}, \bibinfo {author} {\bibfnamefont {T.~X.}\ \bibnamefont
  {Zhou}}, \bibinfo {author} {\bibfnamefont {Q.}~\bibnamefont {Li}}, \bibinfo
  {author} {\bibfnamefont {Y.~J.}\ \bibnamefont {Shin}}, \bibinfo {author}
  {\bibfnamefont {J.~K.}\ \bibnamefont {Shi}}, \bibinfo {author} {\bibfnamefont
  {C.}~\bibnamefont {Burch}}, \bibinfo {author} {\bibfnamefont
  {H.}~\bibnamefont {Zhang}}, \bibinfo {author} {\bibfnamefont
  {F.}~\bibnamefont {Casola}}, \bibinfo {author} {\bibfnamefont
  {T.}~\bibnamefont {Taniguchi}}, \bibinfo {author} {\bibfnamefont
  {K.}~\bibnamefont {Watanabe}}, \bibinfo {author} {\bibfnamefont
  {P.}~\bibnamefont {Kim}}, \bibinfo {author} {\bibfnamefont {A.}~\bibnamefont
  {Yacoby}}, \ and\ \bibinfo {author} {\bibfnamefont {R.~L.}\ \bibnamefont
  {Walsworth}},\ }\href@noop {} {\ }\Eprint
  {http://arxiv.org/abs/http://arxiv.org/abs/1905.10791v1}
  {http://arxiv.org/abs/1905.10791v1} \BibitemShut {NoStop}\bibitem [{\citenamefont {Moll}\ \emph {et~al.}(2016)\citenamefont {Moll},
  \citenamefont {Kushwaha}, \citenamefont {Nandi}, \citenamefont {Schmidt},\
  and\ \citenamefont {Mackenzie}}]{Moll2016}  \BibitemOpen
  \bibfield  {author} {\bibinfo {author} {\bibfnamefont {P.~J.~W.}\
  \bibnamefont {Moll}}, \bibinfo {author} {\bibfnamefont {P.}~\bibnamefont
  {Kushwaha}}, \bibinfo {author} {\bibfnamefont {N.}~\bibnamefont {Nandi}},
  \bibinfo {author} {\bibfnamefont {B.}~\bibnamefont {Schmidt}}, \ and\
  \bibinfo {author} {\bibfnamefont {A.~P.}\ \bibnamefont {Mackenzie}},\ }\href
  {\doibase 10.1126/science.aac8385} {\bibfield  {journal} {\bibinfo  {journal}
  {Science}\ }\textbf {\bibinfo {volume} {351}},\ \bibinfo {pages} {1061}
  (\bibinfo {year} {2016})}\BibitemShut {NoStop}\bibitem [{\citenamefont {Stokes}(1966)}]{Stokes1966}  \BibitemOpen
  \bibfield  {author} {\bibinfo {author} {\bibfnamefont {V.~K.}\ \bibnamefont
  {Stokes}},\ }\href {\doibase 10.1063/1.1761925} {\bibfield  {journal}
  {\bibinfo  {journal} {Physics of Fluids}\ }\textbf {\bibinfo {volume} {9}},\
  \bibinfo {pages} {1709} (\bibinfo {year} {1966})}\BibitemShut {NoStop}\bibitem [{\citenamefont {Mendoza}\ \emph {et~al.}(2013)\citenamefont
  {Mendoza}, \citenamefont {Herrmann},\ and\ \citenamefont
  {Succi}}]{Mendoza2013}  \BibitemOpen
  \bibfield  {author} {\bibinfo {author} {\bibfnamefont {M.}~\bibnamefont
  {Mendoza}}, \bibinfo {author} {\bibfnamefont {H.~J.}\ \bibnamefont
  {Herrmann}}, \ and\ \bibinfo {author} {\bibfnamefont {S.}~\bibnamefont
  {Succi}},\ }\href {\doibase 10.1038/srep01052} {\bibfield  {journal}
  {\bibinfo  {journal} {Scientific Reports}\ }\textbf {\bibinfo {volume} {3}}
  (\bibinfo {year} {2013}),\ 10.1038/srep01052}\BibitemShut {NoStop}\bibitem [{\citenamefont {{Landau}}\ and\ \citenamefont
  {{Lifshitz}}(1959)}]{1959flme.book.....L}  \BibitemOpen
  \bibfield  {author} {\bibinfo {author} {\bibfnamefont {L.~D.}\ \bibnamefont
  {{Landau}}}\ and\ \bibinfo {author} {\bibfnamefont {E.~M.}\ \bibnamefont
  {{Lifshitz}}},\ }\href@noop {} {\emph {\bibinfo {title} {{Fluid
  mechanics}}}}\ (\bibinfo {year} {1959})\BibitemShut {NoStop}\bibitem [{\citenamefont {{Steinberg}}(1958)}]{1958PhRv..109.1486S}  \BibitemOpen
  \bibfield  {author} {\bibinfo {author} {\bibfnamefont {M.~S.}\ \bibnamefont
  {{Steinberg}}},\ }\href {\doibase 10.1103/PhysRev.109.1486} {\bibfield
  {journal} {\bibinfo  {journal} {Physical Review}\ }\textbf {\bibinfo {volume}
  {109}},\ \bibinfo {pages} {1486} (\bibinfo {year} {1958})}\BibitemShut
  {NoStop}\bibitem [{\citenamefont {Neumann}(1885)}]{Neumann1885}  \BibitemOpen
  \bibfield  {author} {\bibinfo {author} {\bibfnamefont {F.}~\bibnamefont
  {Neumann}},\ }\href@noop {} {\bibfield  {journal} {\bibinfo  {journal} {B. G.
  Teubner-Verlag}\ } (\bibinfo {year} {1885})}\BibitemShut {NoStop}\bibitem [{\citenamefont {Nye}(1985)}]{Nye1985}  \BibitemOpen
  \bibfield  {author} {\bibinfo {author} {\bibfnamefont {J.~F.}\ \bibnamefont
  {Nye}},\ }\href
  {https://www.ebook.de/de/product/3237752/j_f_nye_physical_properties_of_crystals.html}
  {\emph {\bibinfo {title} {Physical Properties of Crystals}}}\ (\bibinfo
  {publisher} {Oxford University Press},\ \bibinfo {year} {1985})\BibitemShut
  {NoStop}\bibitem [{Note1()}]{Note1}  \BibitemOpen
  \bibinfo {note} {The viscosity tensors used in each case are provided in
  Supplementary Material.}\BibitemShut {Stop}\bibitem [{\citenamefont {Tomadin}\ \emph {et~al.}(2014)\citenamefont
  {Tomadin}, \citenamefont {Vignale},\ and\ \citenamefont
  {Polini}}]{Tomadin2014}  \BibitemOpen
  \bibfield  {author} {\bibinfo {author} {\bibfnamefont {A.}~\bibnamefont
  {Tomadin}}, \bibinfo {author} {\bibfnamefont {G.}~\bibnamefont {Vignale}}, \
  and\ \bibinfo {author} {\bibfnamefont {M.}~\bibnamefont {Polini}},\ }\href
  {\doibase 10.1103/physrevlett.113.235901} {\bibfield  {journal} {\bibinfo
  {journal} {Physical Review Letters}\ }\textbf {\bibinfo {volume} {113}}
  (\bibinfo {year} {2014}),\ 10.1103/physrevlett.113.235901}\BibitemShut
  {NoStop}\bibitem [{\citenamefont {Ella}\ \emph {et~al.}(2019)\citenamefont {Ella},
  \citenamefont {Rozen}, \citenamefont {Birkbeck}, \citenamefont {Ben-Shalom},
  \citenamefont {Perello}, \citenamefont {Zultak}, \citenamefont {Taniguchi},
  \citenamefont {Watanabe}, \citenamefont {Geim}, \citenamefont {Ilani},\ and\
  \citenamefont {Sulpizio}}]{Ella2019}  \BibitemOpen
  \bibfield  {author} {\bibinfo {author} {\bibfnamefont {L.}~\bibnamefont
  {Ella}}, \bibinfo {author} {\bibfnamefont {A.}~\bibnamefont {Rozen}},
  \bibinfo {author} {\bibfnamefont {J.}~\bibnamefont {Birkbeck}}, \bibinfo
  {author} {\bibfnamefont {M.}~\bibnamefont {Ben-Shalom}}, \bibinfo {author}
  {\bibfnamefont {D.}~\bibnamefont {Perello}}, \bibinfo {author} {\bibfnamefont
  {J.}~\bibnamefont {Zultak}}, \bibinfo {author} {\bibfnamefont
  {T.}~\bibnamefont {Taniguchi}}, \bibinfo {author} {\bibfnamefont
  {K.}~\bibnamefont {Watanabe}}, \bibinfo {author} {\bibfnamefont {A.~K.}\
  \bibnamefont {Geim}}, \bibinfo {author} {\bibfnamefont {S.}~\bibnamefont
  {Ilani}}, \ and\ \bibinfo {author} {\bibfnamefont {J.~A.}\ \bibnamefont
  {Sulpizio}},\ }\href {\doibase 10.1038/s41565-019-0398-x} {\bibfield
  {journal} {\bibinfo  {journal} {Nature Nanotechnology}\ }\textbf {\bibinfo
  {volume} {14}},\ \bibinfo {pages} {480} (\bibinfo {year} {2019})}\BibitemShut
  {NoStop}\end{thebibliography}
\end{document}